\documentclass{iopjournal}

\usepackage{amsmath, bm, amssymb}
\usepackage{mathptmx}
\usepackage[cmintegrals,varg,bigdelims]{newtxmath}

\usepackage{doi}
\usepackage{cite}
\usepackage{mathtools}  
\usepackage[mathlines]{lineno}
\usepackage{etoolbox}

\usepackage{fancyhdr}
\renewcommand{\marginpar}[1]{}

\begin{document}

\title{
A homoclinic route to chaos in omnivore communities}

\author{
	Yiyuan Niu$^{1,\dagger}$,
	Ju Kang$^{2,\dagger,\ddagger}$, 
	Wei Tao$^1$,
	Xin Wang$^{1,*}$
}

\affil{$^1$School of Physics, Sun Yat-sen University, Guangzhou 510275, China}\\
\affil{$^2$School of Ecology, Sun Yat-sen University, Shenzhen 518107, China}\\
\affil{$\dagger$ These authors contributed equally to this work}\\		
\affil{$\ddagger$ Corresponding author: \href{mailto:kangj29@mail.sysu.edu.cn}{kangj29@mail.sysu.edu.cn}}\\
\affil{* Corresponding author: \href{mailto:wangxin36@mail.sysu.edu.cn}{wangxin36@mail.sysu.edu.cn}}\\

\keywords{Omnivory, intraguild predation, Shilnikov homoclinic orbit, chaos}

\begin{abstract}
Omnivory, where species feed across multiple trophic levels, is a widespread feature of ecological networks. 
A key mechanism underlying such complexity is intraguild predation (IGP), 
in which a top predator consumes both an intermediate predator and a shared resource. 
Here, we show that Shilnikov homoclinic orbits emerge in a minimal intraguild predation model, 
triggering a cascade of homoclinic bifurcations near a saddle-focus equilibrium 
that culminates in chaos. 
Numerical simulations and Lyapunov spectrum analysis reveal multiple coexistence modes, 
ranging from regular oscillations to Shilnikov homoclinic orbits and chaos.
Our model quantitatively reproduces patterns observed in natural omnivore networks,
providing mechanistic insights into complex population fluctuations in ecological systems.
\end{abstract}

\section*{1 \quad Introduction}
\stepcounter{section}
\label{intro}
Understanding the mechanisms that enable complex, self-organized population dynamics in natural systems remains a central focus of theoretical ecology~\cite{
	Costantino1995,Blasius1999, JuChu2025}.
Omnivory, in which species feed across multiple trophic levels, is a widespread feature of both aquatic and terrestrial ecosystems~\cite{
	Cousins1987, PolisStrong1996, Thompson2007}, 
generating cross-trophic interactions that can profoundly influence population fluctuations, community persistence, and ecosystem stability~\cite{Pimm1978}. 
A key form of omnivory is intraguild predation (IGP), 
in which a top predator consumes both a basal resource and a competing intermediate consumer. 
By combining competition and predation within a single trophic module, 
IGP systems can exhibit complex dynamical behaviors~\cite{
	Kratina2012,Hiltunen2013,BanerjeeSen2018}.
Previous studies have shown that IGP can stabilize populations through mechanisms 
such as adaptive foraging and morph switching~\cite{
	Arim2004, Langley2025, McCannHasting1997, WoodieAnderson2024}, 
yet it can also destabilize them, producing large-amplitude oscillations or 
chaotic fluctuations~\cite{Cicero2024, Dash2024, Han2024, Su2025, Farivar2025}.

In IGP systems, chaotic population dynamics are often attributed to period-doubling 
bifurcations~\cite{TanabeNumba2005,MoitriSen2012}.
Yet, other mechanisms capable of generating chaos remain largely unexplored. 
One such mechanism is the Shilnikov homoclinic bifurcation, 
in which a trajectory departing from a saddle-focus equilibrium returns along its stable manifold, 
forming a homoclinic loop that gives rise to a Smale horseshoe structure 
and a countable set of periodic orbits, leading to chaotic dynamics~\cite{
	Shilnikov1972, Kuznetsov2001}.
While Shilnikov-type chaos has been extensively studied in 
simple food-chain models~\cite{
	DengHines2002, WuNi2024, McCann1995, HastingsPowell1991},
its potential role in IGP systems has not been examined.

Here, we develop a minimal model of intraguild predation~\cite{Holt1989, HoltPolis1997}, 
and demonstrate that Shilnikov-type homoclinic orbits and chaotic dynamics 
can arise in IGP systems, giving rise to complex, self-organized population fluctuations. 
By combining numerical simulations, local bifurcation analysis, 
and numerical evaluation of Lyapunov exponents, 
we identify the conditions under which distinct coexistence modes emerge independently of 
stochastic forcing or environmental variability. 
The model is applicable to natural IGP communities. 
In particular, our results show that the simulated self-organized oscillations quantitatively 
reproduce empirical field data from a host-parasite system~\cite{Borer2003}
across communities with differing productivity levels. 
Overall, this study provides new insights into how intraguild predation 
can drive the emergence of complex, self-organized dynamics in natural ecosystems.

\section*{2 \quad Results}
\stepcounter{section}
\label{sec2results}
\subsection*{2.1 \quad A minimal model of intraguild predation}
\stepcounter{subsection}
\label{model}

\begin{figure}[ht!]
\centering
\includegraphics[width=10cm]{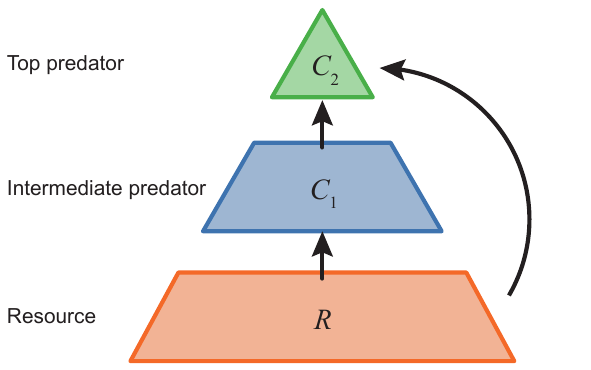}
\caption{\label{illustration} 
	Schematic of a minimal intraguild predation model.
	An intermediate predator $C_1$ feeds on the basal resource $R$, 
	while the top predator $C_2$ preys upon both $R$ and $C_1$, 
	generating a dual interaction of competition and predation. 
	Arrows indicate the direction of biomass flow.}
\end{figure}

To investigate the effects of omnivory on complex population dynamics 
in simple communities,
we developed a minimal model of intraguild predation (Eq.~\ref{P-Psystem1}) 
involving three interacting populations: a basal resource $R$, 
an intermediate predator $C_1$, 
and a top predator $C_2$ (Fig.~\ref{illustration}). 
This three-species configuration represents the simplest ecological framework 
in which both predation and competition coexist.
Feeding interactions follow the classical Holling type-II functional response~\cite{HollingCS1965},
which describes saturation effects in consumer feeding rates commonly observed 
across aquatic and terrestrial systems~\cite{Abrams2022}. 

\noindent The population dynamics of the system can thus be described 
by the following equations:
\begin{equation}
\begin{dcases}
	\frac{dR}{dt} 
&= rR\!\left(1 - \frac{R}{K_0}\right)
 - \frac{a_{1,R}\,R}{1 + b_{1,R}\,R}\; C_1
 - \frac{a_{2,R}\,R}{1 + b_{2,R}\,R}\; C_2, \\[6pt]
\frac{dC_1}{dt} 
&= w_{1,R}\,\frac{a_{1,R}\,R}{1 + b_{1,R}\,R}\; C_1
 - \frac{a_{2,C_1}\,C_1}{1 + b_{2,C_1}\,C_1}\; C_2
 - m_{1}\, C_1, \\[6pt]
\frac{dC_2}{dt} 
&= w_{2,R}\,\frac{a_{2,R}\,R}{1 + b_{2,R}\,R}\; C_2
 + w_{2,C_1}\,\frac{a_{2,C_1}\,C_1}{1 + b_{2,C_1}\,C_1}\; C_2
 - m_{2}\, C_2.
\end{dcases}
\label{P-Psystem1}
\end{equation}
Here, $r$ and $K_0$ denote the intrinsic growth rate and carrying capacity 
of the basal biotic resource.
$a_{i,s}$ and $\ensuremath{1/b_{i,s}}$ represent the attack rate and 
saturation coefficient characterizing consumer $i$'s functional response 
to prey species $s$, where $s \in \left\{R, C_1\right\}$ 
and $i \in \left\{1, 2\right\}$. 
The term $w_{i,s}$ denotes the biomass conversion efficiency, 
and $m_i$ represents the mortality rates of consumer $i$.
All state variables remain non-negative under biologically feasible conditions.

\subsection*{2.2 \quad Emergence of a Shilnikov homoclinic orbit}
\stepcounter{subsection}
\label{Shilnikov}
The IGP system described by Eq.~(\ref{P-Psystem1}) admits five biologically relevant equilibria.
The null state, $E_0 = \left(0, 0, 0\right)$, 
corresponds to the extinction of all populations.
The basal-resource state, $E_R = \left(K_0, 0, 0\right)$, 
represents the condition where only the resource persists at its carrying capacity.
Two single-consumer equilibria also exist,
$E_1 = \left(\rho_1, \xi_1, 0 \right)$ and 
$E_2 = \left(\rho_2, 0, \xi_2\right)$, where
$\rho_1 = \frac{m_1}{w_{1,R}\,a_{1,R}-b_{1,R}\,m_1}$, 
$\xi_1 = \rho_1\frac{r\,w_{1,R}}{m_1}\left(1 - \frac{\rho_1}{K_0}\right)$, 
$\rho_2 = \frac{m_2}{w_{2,R}\,a_{2,R}-b_{2,R}\,m_2}$,
$\xi_2 = \rho_2\frac{r\,w_{2,R}}{m_2}\left(1 - \frac{\rho_2}{K_0}\right)$,
with $C_1$ and $C_2$ coexisting with the resource, respectively.
Finally, the coexistence equilibrium, $E^{*} = \left(R^{*}, C_1^{*}, C_2^{*}\right)$,
represents the state in which all three populations persist simultaneously.

\begin{figure}[ht!]    
	\centering
	\includegraphics[width=14cm]{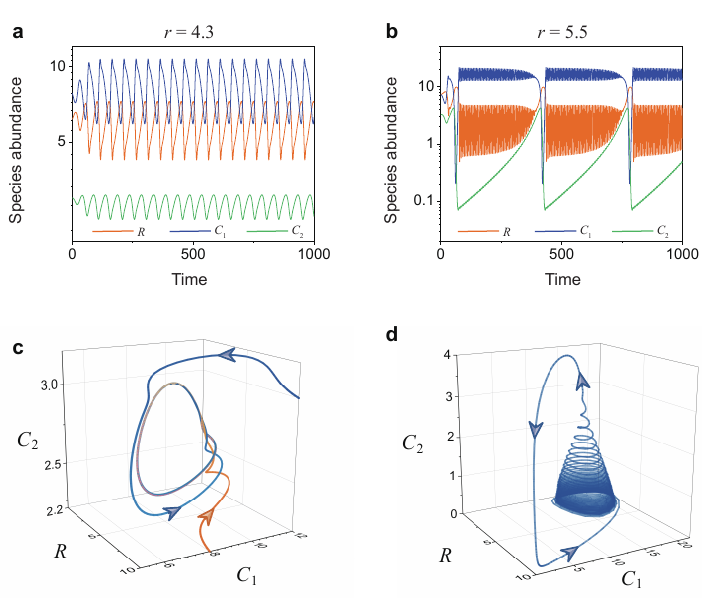}
	\caption{\label{HomoclinicChaos} 
Emergence of Shilnikov homoclinic attractors in a minimal intraguild predation model
as resource productivity $r$ increases across different ecosystems.
	(a-b) Time series of population abundances, showing the transition 
	from regular oscillations to Shilnikov homoclinic oscillations.
	(c-d) The corresponding 3D phase-space trajectories associated with panel (a-b), 
	demonstrating the transition from a stable limit cycle to a homoclinic orbit.
	See SM Sec. III for simulation details of Figs.~\ref{HomoclinicChaos}-\ref{CompareExp}.} 
\end{figure}

Next, we assess the local stability of the coexistence equilibrium $E^{*}$
by analyzing the eigenvalues of its Jacobian matrix $J\left(E^{*}\right)$.
The eigenvalues $\left(\lambda_1, \lambda_2, \lambda_3\right)$ 
satisfy the characteristic polynomial:
$\lambda^3 + \sigma_1 \lambda^2 + \sigma_2 \lambda + \sigma_3 = 0$,
where $\sigma_1, \sigma_2$, and $\sigma_3$ are real coefficients
determined by the system parameters.
According to the Routh-Hurwitz criterion, $E^{*}$ is locally stable 
if $\sigma_1, \sigma_2 > 0$ and $\sigma_1 \sigma_2 > \sigma_3$.
Violation of any of these conditions destabilizes the equilibrium, 
potentially leading to a saddle-node or Hopf bifurcation.
A saddle-focus equilibrium arises when $J\left(E^{*}\right)$ 
has one real eigenvalue $\lambda_1 \in \mathbb{R}$ and
a complex conjugate pair $\lambda_{2,3} = \mu \pm i \delta $, 
with $\mu, \delta \in \mathbb{R}$ and $\delta \neq 0$,
where $\mu$ and $\delta$ denote the real and imaginary parts, respectively.
A saddle-focus satisfying the Shilnikov conditions 
$\mu \, \lambda_1 < 0$ and $|\mu| < |\lambda_1|$
is a prerequisite for a Shilnikov homoclinic orbit, 
a classical route to chaos. 

Consistent with these criteria, numerical simulations confirm 
$E^{*} = \left(7.35, 7.36, 3.29\right)$ as a saddle-focus.
The corresponding eigenvalues are $\lambda_1 = -3.03$
and $\lambda_{2,3} = 0.108 \pm 0.098 i$, satisfying the Shilnikov conditions
for a homoclinic orbit.
Trajectories diverge from $E^{*}$ 
and return along its stable manifold, 
forming a closed homoclinic loop, consistent with a Shilnikov homoclinic orbit~\cite{
	QiongHe2016} (Fig.~\ref{HomoclinicChaos}b).
Increasing resource productivity $r$
induces a transition from limit-cycle oscillations 
to a Shilnikov homoclinic orbit (Fig.~\ref{HomoclinicChaos}c-d).
These shifts indicate that resource enrichment destabilizes 
the coexistence equilibrium, 
generating self-sustained oscillation via the Shilnikov mechanism. 

\subsection*{2.3 \quad Shilnikov-type chaos through homoclinic bifurcations}
\stepcounter{subsection}
\label{strange}

\begin{figure}[ht!]
    \centering
    \includegraphics[width=14cm]{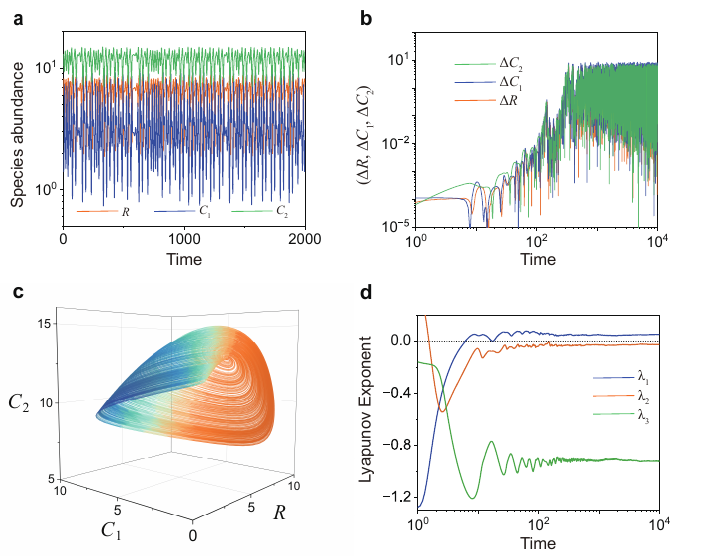}
    \caption{\label{Smale}
    Chaotic dynamics in a minimal intraguild predation model.
    (a) Time series of predator and prey abundances showing irregular, 
	aperiodic oscillations.
    (b) Divergence of trajectories under small perturbations ($\Delta C_{1}=10^{-4}$), 
	demonstrating sensitivity to initial conditions.
    (c) Chaotic attractor in the $\left(R,C_1,C_2\right)$ phase space, 
	showing horseshoe-type strange attractor.
    (d) Temporal evolution of the Lyapunov exponents computed 
	by the Benettin algorithm; 
	the positive $\lambda_{1}$ and negative sum of exponents confirm chaos.
	}
\end{figure}

A Shilnikov-type homoclinic bifurcation 
can induce chaos near
the saddle-focus equilibrium $E^{*}$.
To characterize the chaotic dynamics of the IGP system,
we performed sensitivity analysis and computed the Lyapunov spectrum (Fig.~\ref{Smale}).
The prey ($R$) and predators ($C_1$, $C_2$) exhibit aperiodic, irregular oscillations,
providing direct evidence of chaos (Fig.~\ref{Smale}a).
A small perturbation in the initial predator abundance ($\Delta C_{1}=10^{-4}$) 
causes trajectories to diverge rapidly (Fig.~\ref{Smale}b), 
demonstrating sensitivity to initial conditions, 
a hallmark of chaotic dynamics.
In the three-dimensional phase space $\left(R, C_1, C_2\right)$,
a single trajectory traces a horseshoe-shaped strange attractor (Fig.~\ref{Smale}c),
consistent with Shilnikov-type chaos.
Lyapunov exponents reveal a positive largest exponent ($\lambda_1 = 0.05 > 0$),
and a negative sum of all exponents ($\Sigma_i \lambda_i = -0.89 < 0$),
confirming exponential divergence of nearby trajectories 
and overall dissipative behavior (Fig.~\ref{Smale}d).
These spectral properties indicate chaos confined 
to a bounded strange attractor.

The saddle-focus $E^{*}$ serves as the organizing center of the chaotic dynamics.
Homoclinic bifurcations destabilize nearby orbits, 
producing recurrent, irregular fluctuations across trophic levels 
while all three species coexist. 
These results demonstrate that even a minimal intraguild predation model
can generate chaos via a Shilnikov-type mechanism, 
revealing a novel pathway for complex population dynamics induced by omnivory.

\subsection*{2.4 \quad Comparison of model predictions with field observations}
\stepcounter{subsection}
\label{compare_observation}

\begin{figure}[ht!]
    \centering
    \includegraphics[width=14cm]{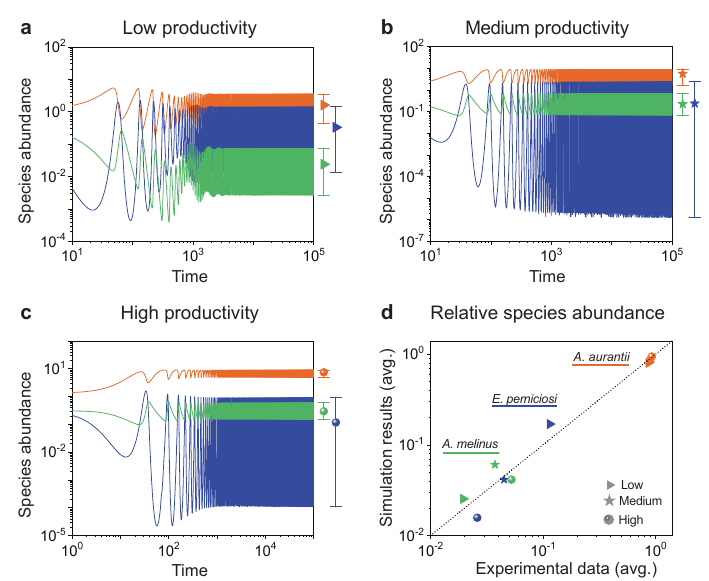}
    \caption{\label{CompareExp}
    Comparison of model results with field data from a 
	host-parasite community~\cite{Borer2003}.
    (a-c) Simulated time series of species abundances under 
	increasing resource productivity (modeled via parameter $r$).
    Markers indicate time-averaged abundances, 
	and error bars represent oscillation amplitudes.
    (d) Comparison of model results and experimental data~\cite{Borer2003}
	based on time-averaged relative abundances across productivity gradients 
	(grapefruit: low; citrus: medium; lemon: high).
 	Model-experiment correspondence is quantified using Bray-Curtis similarity, 
	with values of 0.93 (low), 0.97 (medium) and 0.97 (high).
	Species are color-coded as follows:
	$R$ in orange, $C_1$ in blue, and $C_2$ in green.}
\end{figure}

We evaluate the model against field data from a host-parasite community~\cite{Borer2003},
a representative IGP system.
This community includes the specialist parasitoid \textit{Encarsia perniciosi} 
(intermediate predator)
and the facultative parasitoid \textit{Aphytis melinus} (top predator), 
both exploiting the common host \textit{Aonidiella aurantii} 
(California red scale)~\cite{Borer2003}.
Model dynamics are examined across three productivity levels 
(grapefruit: low, citrus: medium, and lemon: high) by varying 
the resource growth rate ($r$), 
reflecting observed differences in host egg densities among cultivars~\cite{
	Hare1990, Borer2003}.

Across all levels, communities exhibit sustained self-organized oscillations 
	(Fig.~\ref{CompareExp}a-c).
At low productivity, the intermediate predator ($C_1$) dominates,
whereas high productivity strengthens direct resource-omnivore interactions, 
elevating the mean abundance of the top predator ($C_2$).
Meanwhile, the time-averaged relative abundance of the basal resource ($R$) 
increases from 0.87 to 0.92, 
highlighting productivity-dependent shifts in species dominance.

Furthermore, we quantify model performance using Bray-Curtis similarities 
between simulated and observed time-averaged relative abundances,
yielding values above 0.9 across all three productivity levels (Fig.~\ref{CompareExp}d),
exceeding the widely accepted ecological threshold of 0.8~\cite{
	Bray1957}.
This strong correspondence demonstrates that 
the model reliably captures community-level patterns
and indicates that omnivory interactions within the intraguild 
predation framework are sufficient to generate ecologically realistic dynamics.

\section*{3 \quad Discussion} 
\stepcounter{section}
\label{sec5}
Omnivory through intraguild predation is pervasive in natural ecosystems 
and has been recognized as a key driver of complex population dynamics~\cite{
	Polis1991,HoltPolis1997,Hiltunen2013}.
Although mechanisms such as adaptive foraging and morphological switching 
that stabilize IGP systems have been extensively studied~\cite{
	KrivanDiehl2005,Mylius2001}, 
the role of Shilnikov homoclinic bifurcations in shaping omnivory dynamics 
remains largely unexplored. 
Here, we show that a minimal intraguild predation model can 
facilitate Shilnikov homoclinic bifurcations 
and give rise to intrinsic chaotic dynamics near a saddle-focus equilibrium. 
By combining numerical simulations and Lyapunov spectrum analysis, 
we reveal self-organized coexistence modes ranging from 
regular oscillations to Shilnikov homoclinic orbits 
and Shilnikov-type chaotic fluctuations. 
Furthermore, our model quantitatively explains the species coexistence patterns of 
a host-parasite community, which constitutes a natural IGP community 
observed in field studies~\cite{Borer2003}. 

Previous studies of IGP systems have emphasized the route to chaos 
via period-doubling bifurcations~\cite{Pal2014, TanabeNumba2005, MoitriSen2012}.
In contrast, our findings reveal Shilnikov-type bifurcations 
leading to chaos in IGP systems. 
This mechanism involves a global bifurcation 
in which trajectories depart from a saddle-focus coexistence equilibrium 
and return along its stable manifold, forming a homoclinic loop that 
generates a Smale horseshoe structure and a countable set of 
unstable periodic orbits~\cite{Shilnikov1972, Kuznetsov1995, Kuznetsov2001}.
It produces irregular bursting, long excursions near the equilibrium, 
and sensitive dependence on initial conditions, 
yielding transient and spectral signatures distinct from 
those of period-doubling cascades. 
Although Shilnikov-type chaos has been reported in simple food-chain systems 
lacking omnivory~\cite{DengHines2002, WuNi2024, McCann1995, HastingsPowell1991},
omnivory and intraguild predation are widespread features 
of natural ecosystems~\cite{Cousins1987}.
Therefore, our results highlight the ecological relevance 
of Shilnikov-type dynamics in IGP systems 
and provide a mechanistic explanation for intrinsic population fluctuations 
observed in field data. 

Finally, from an ecological perspective, 
intraguild predation plays a key role in promoting species coexistence~\cite{Brose2019}.
A well-known constraint on species diversity in natural ecosystems 
is the competitive exclusion principle (CEP), 
which states that two consumer species competing for a single type of resource 
cannot coexist at steady state~\cite{G.F.Gause1934, GarrettHardin1960, 
	RobertMacArthur1964, SimonA.Levin1970, RichardMcGehee1977}.
Although our previous studies have identified mechanisms that can break CEP, 
such as pack hunting~\cite{Wangxin2020} and intraspecific predation~\cite{JuKang2024,JuKangCSF}, 
intraguild predation promotes species coexistence through 
a food-web mechanism that alleviates the constraints of CEP 
by creating contextual differences~\cite{JuKangQB}.
Disruption of omnivory, whether due to behavioral or environmental constraints 
in an IGP system, can impair species coexistence that would otherwise 
be facilitated (Fig.~S1). 
Our results provide an explanation for sustained species coexistence 
under resource enrichment and offer new insights 
into the paradox of enrichment~\cite{Rozenweig1971}.

\section*{Acknowledgments}
	We thank Fan Zhong for helpful discussions. This work was supported by National Natural Science Foundation of China (Grant No. 12474207). 
	
\section*{Conflict of interest}
The authors declare that they have no known competing financial interests or personal relationships that could have appeared to influence the work reported in this paper.

 	\section*{Author contributions}
	J.K. and X.W. conceived the project and planned the study. All authors developed the model, performed the theoretical analysis and numerical simulations, analyzed the data, and wrote the manuscript.
	\section*{Data and Code Availability}
	All data and code used in this study are available from the corresponding authors upon reasonable request.	

	\bibliography{reference.bib}

\begin{thebibliography}{55}

	\bibitem{QiongHe2016} Qiong He, Hai-Yun Xiong. Shilnikov chaos and Hopf bifurcation in three-dimensional differential system. \href{https://dx.doi.org/10.1016/j.ijleo.2016.05.149}{Optik 2016; 127(19): 7648-7655}.	
		
\end{thebibliography}
	
\newpage
\clearpage

\pagestyle{plain}
\fancyfoot[C]{\thepage}

\renewcommand{\thefigure}{S\arabic{figure}}
\renewcommand{\theequation}{S\arabic{equation}}
\renewcommand{\thetable}{S\arabic{table}}
\renewcommand{\thesection}{\Roman{section}}
\renewcommand{\thesubsection}{\Alph{subsection}}
\renewcommand{\thesubsubsection}{\arabic{subsubsection}}

\setcounter{page}{1}
\setcounter{section}{0}
\setcounter{subsection}{0}
\setcounter{figure}{0}
\setcounter{table}{0}
\setcounter{equation}{0}


\begin{center}
Supplementary Material for\\
\textbf{\large{
A homoclinic route to chaos in omnivore communities}}
\end{center}

\begin{center}
	Yiyuan Niu, Ju Kang, Wei Tao, Xin Wang\\
\normalsize{Corresponding author. E-mail: kangj29@mail.sysu.edu.cn; wangxin36@mail.sysu.edu.cn}
\end{center}
\vspace{2em}

\tableofcontents

\newpage

\section{Stability analysis of coexistent equilibrium}\label{Stability analysis}

To facilitate the stability analysis, we first nondimensionalize the minimal intraguild predation model (see Eq.~(1))
using the substitutions
\[
\tilde{R}=\dfrac{R}{K_0}, \quad \tilde{C_1}=a_{1,R}C_1, \quad \tilde{C_2}=a_{2,R}C_2.
\]
The system then depends on the dimensionless variables $(\tilde{R}, \tilde{C}_1, \tilde{C}_2)$. 
For notational simplicity, we drop the tildes writing $(R, C_1, C_2)$
for the dimensionless variables. 
Equivalently, this transformation corresponds to:
\[
R \mapsto K_0 R, \quad C_1 \mapsto \dfrac{C_1}{a_{1,R}}, \quad C_2 \mapsto \dfrac{C_2}{a_{2,R}}.
\]
Under this convention, the dimensionless system takes the form:
\begin{equation}
\begin{cases}
\dfrac{dR}{dt}=rR(1-R)-\dfrac{RC_{1}}{\alpha_{1}R+1}-\dfrac{RC_{2}}{\alpha_{2}R+1} \equiv g_1, \\[8pt] 
\dfrac{dC_{1}}{dt}=\dfrac{\omega_{1}RC_{1}}{\alpha_{1}R+1}
-\dfrac{\alpha_{4}C_{1}C_{2}}{\alpha_{3}C_{1}+1}-m_1C_{1} \equiv g_2, \\[8pt] 
\dfrac{dC_{2}}{dt}=\dfrac{\omega_{2}RC_{2}}{\alpha_{2}R+1}
+\dfrac{\omega_{3}C_{1}C_{2}}{\alpha_{3}C_{1}+1}-m_2C_{2} \equiv g_3.
\end{cases}
\label{P-Psystem2}
\end{equation}
The corresponding dimensionless parameters are defined as:
\[
\alpha_i = b_{i,R} K_0 \, (i=1,2), \quad \alpha_3 = \frac{b_{2,C_1}}{a_{1,R}}, \quad \alpha_4 = \frac{a_{2,C_1}}{a_{2,R}}, \quad \omega_1 = w_{1,R} a_{1,R} K_0, \quad \omega_2 = w_2 a_{2,R} K_0, \quad \omega_3 = w_{2,C_1} \frac{a_{2,C_1}}{a_{1,R}}.
\]
The parameters $r$,  $\{m_i\}_{i=1}^{2}$, $\{\omega_i\}_{i=1}^{3}$, and $\{\alpha_i\}_{i=1}^{4}$ 
collectively govern the local and global dynamics of Eq.~(\ref{P-Psystem2}),
and determine the stability of the coexistence state.

The equilibria of Eq.~(\ref{P-Psystem2}) can be obtained by solving equations 
$\dot{R}=0,\dot{C_{1}}=0$ and $\dot{C_{2}}=0$. 
Computation shows that Eq.~(\ref{P-Psystem2}) admits a single coexistence equilibrium 
$E^{*}=(R^{*},C_{1}^{*},C_{2}^{*}) > 0$, which satisfies:
\begin{equation}
	\begin{cases}
		r{R}(1-{R})-\dfrac{R{C}_{1}}{\alpha_{1}R+1}-\dfrac{R{C}_{2}}{\alpha_{2}R+1}=0, \\[8pt] 
		\dfrac{\omega_{1}R{C}_{1}}{\alpha_{1}R+1}-\dfrac{\alpha_{4}{C}_{1}C_{2}}{\alpha_{3}{C}_{1}+1}-m_1{C}_{1}=0, \\[8pt] 
		\dfrac{\omega_{2}R{C}_{2}}{\alpha_{2}R+1}+\dfrac{\omega_{3}{C}_{1}C_{2}}{\alpha_{3}{C}_{1}+1}-m_2{C}_{2}=0.  \\[8pt] 
	\end{cases}
	\label{coexistingequilibrium}
\end{equation}
The equation also possesses four extinction equilibria: \\
	$E_{0}=(0,0,0)$, $E_{R}=(1,0,0)$, 
	$E_{1}=\left( \frac{m_1}{\omega_1 - \alpha_1 m_1},\frac{\omega_1 (\omega_1 - (1 + \alpha_1) m_1) r}{(\omega_1 - \alpha_1 m_1)^2},0\right) $ 
	and	$E_{2}=\left( \frac{m_2}{\omega_2 - \alpha_2 m_2},0,\frac{\omega_2 (\omega_2 - (1 + \alpha_2) m_2) r}{(\omega_2 - \alpha_2 m_2)^2}\right) $.

\noindent The Jacobian matrix of the Eq.~(\ref{P-Psystem2}) at $E^{*}$ is:
\begin{equation}
J(E^*) = [b_{ij}] = \left[\dfrac{\partial g_i}{\partial x_j}\right]_{E^*}, \quad (i,j = 1,2,3),
\label{Jacobian}
\end{equation}
where $b_{ij}$ denotes the $\left(i, j\right)$ entry of $J(E^{*})$. 
The eigenvalues $\left(\lambda_1, \lambda_2, \lambda_3\right)$ 
of the Jacobian matrix $J(E^{*})$ satisfy the following characteristic polynomial:
\begin{equation}
	\lambda^{3}+\sigma_{2}\lambda^{2}+\sigma_{1}\lambda+\sigma_{0}=0.
	\label{eigenequation}
\end{equation}
with\quad $\sigma_{2}=\sum\limits_{i=1}^{3}b_{ii}$,\qquad
$\sigma_{1}=b_{12}b_{21}+b_{13}b_{31}+b_{23}b_{32}-b_{22}b_{33}-b_{11}(b_{22}+b_{33})$, \\
and \quad $\sigma_{0}=b_{12}b_{23}b_{31}+b_{13}b_{21}b_{32}-b_{11}b_{23}b_{32}-b_{12}b_{21}b_{33}+b_{11}b_{22}b_{33}-b_{13}b_{22}b_{31}$.

\noindent Define 
\begin{equation}
	\sigma_{3}=\sigma_{1}\sigma_{2}-\sigma_{0}.
	\label{Routh}
\end{equation}

According to the Routh-Hurwitz criterion, the coexistent equilibrium point $E^*$ 
is locally asymptotically stable if $\sigma_0 > 0$, $\sigma_2 > 0$, and $\sigma_1 \sigma_2 > \sigma_3$ simultaneously. 
Otherwise, $E^*$ is unstable.

\section{Homoclinic orbit analysis}\label{Homoclinic orbit analysis}
For the coexistence equilibrium $E^{*}$ to be a candidate for the Shilnikov bifurcation,
the eigenvalues must $\left(\lambda_1, \lambda_2, \lambda_3\right)$ satisfy Shilnikov conditions
($\mu$ and $\delta$ denote the real and imaginary parts of $\lambda_{2, 3}$, respectively): 
\begin{itemize}
	\item One real eigenvalue: $\lambda_1 \in \mathbb{R}$,
	\item One complex conjugate pair: $\lambda_{2,3} = \mu \pm i \delta $, where $\mu, \delta \in \mathbb{R}$ and $\delta \neq 0$,
	\item  $\mu \, \lambda_1 < 0$ and $|\mu| < |\lambda_1|$.
\end{itemize}
Under these conditions, the system exhibits a Shilnikov homoclinic orbit, 
and the associated bifurcation generates countably many unstable periodic orbits and chaotic invariant sets.

The characteristic polynomial of the Jacobian $J(E^{*})$ (see Eq.~(\ref{eigenequation})) 
can be rewritten into the following form:
\begin{equation}
	\Lambda^{3}+p\Lambda+q=0.
	\label{eigenequation0}
\end{equation}
where 
\begin{equation}
	\lambda=\Lambda-\sigma_{2}/3,\quad p=\sigma_{1}-\sigma_{2}^{2}/3,\quad q=\sigma_{0}-\sigma_{1}\sigma_{2}/3 +2\sigma_{2}^{3}/27.
	\label{Delta0}
\end{equation}
The discriminant ($\Delta$) is defined as
\begin{equation}
	\Delta=\left( \frac{q}{2}\right)^{2}+\left(\frac{p}{3}\right)^{3}.
	\label{Delta}
\end{equation}

When $\Delta > 0$, then Eq.~(\ref{eigenequation0}) has one real root $\Lambda_{1}$ 
and a complex conjugate pair $\Lambda_{2,3}=\gamma \pm i\delta$, where
\begin{equation}
\begin{cases}
\Lambda_1 = \sqrt[3]{-\dfrac{q}{2} + \sqrt{\Delta}} + \sqrt[3]{-\dfrac{q}{2} - \sqrt{\Delta}}, \\[6pt]
\gamma = -\dfrac{1}{2}\left(\sqrt[3]{-\dfrac{q}{2} + \sqrt{\Delta}} + \sqrt[3]{-\dfrac{q}{2} - \sqrt{\Delta}}\right), \\[6pt]
\delta = \dfrac{\sqrt{3}}{2}\left(\sqrt[3]{-\dfrac{q}{2} + \sqrt{\Delta}} - \sqrt[3]{-\dfrac{q}{2} - \sqrt{\Delta}}\right).
\end{cases}
\label{condition}
\end{equation}
The corresponding roots of Eq.~(\ref{eigenequation}) are
\[
\lambda_1 = \Lambda_1 - \frac{\sigma_2}{3}, \quad \lambda_2 = \gamma - \frac{\sigma_2}{3} + i \delta, \quad \lambda_3 = \gamma - \frac{\sigma_2}{3} - i \delta.
\]
The coexistence equilibrium $E^{*}$ is a saddle focus, as indicated by 
$\lambda_{1}\lambda_{2}\lambda_{3}=\lambda_1\left(\mu^2+\delta^2\right)=-\sigma_{0}<0$, with $\lambda_{1}<0$.
The Shilnikov theorem requires the spectral conditions, $ \mu > 0$ and $|\mu|<|\lambda_1|$ to be satisfied.
Given that $\sigma_2 = \sum\limits_{i=1}^{3}b_{ii} < 0$,
which reflects the intraspecific competition in the system (see Eq.~\ref{P-Psystem2}),
these conditions are equivalent to the single inequality:
\begin{equation}
\sqrt[3]{-\dfrac{q}{2} + \sqrt{\Delta}} + \sqrt[3]{-\dfrac{q}{2} - \sqrt{\Delta}} < \dfrac{4 \sigma_2}{3}.
\label{condition1}
\end{equation}
where $ \Delta=\left( \frac{q}{2}\right)^{2}+\left(\frac{p}{3}\right)^{3} $. 

Therefore, if $\Delta>0$ and the inequality~(\ref{condition1}) hold, 
the minimal intraguild predation system (see Eq.~\ref{P-Psystem2}) exhibits a Shilnikov homoclinic orbit~\cite{QiongHe2016}, 
generating countably many unstable periodic orbits and chaotic invariant sets.

\section{Simulation details}\label{Simulation details}

In Fig.2(a-d): 
$K_0 = 10$, 
$w_{1,R} = 0.7$, $a_{1,R} = 0.35$, $b_{1, R} = 0.2$, 
$w_{2,R} = 0.1$, $a_{2,R} = 0.45$, $b_{2,R} = 0.35$,
$w_{2,C_1} = 0.53$, $a_{2,C_1} = 0.67$, $b_{2,C_1} = 0.64$
$m_1 = 0.35$, $m_2=0.55$.
In Fig. 2(a,c): $r = 4.3$. In Fig. 2(b,d): $r= 5.5$.

In Fig.~3(a-d): 
$K_0 = 10$, 
$w_{1,R} = 1$, $a_{1,R} = 0.23$, $b_{1, R} = 0.05$, 
$w_{2,R} = 0.05$, $a_{2,R} = 0.064$, $b_{2,R} = 0.25$,
$w_{2,C_1} = 0.98$, $a_{2,C_1} = 0.14$, $b_{2,C_1} = 0.37$
$m_1 = 0.3$, $m_2=0.2$.

In Fig.~4(a-d):
$K_0 = 10$, 
$w_{1,R} = 0.7$, $a_{1,R} = 0.15$, $b_{1, R} = a_{2,R}/0.7$, 
$w_{2,R} = 0.1$, $a_{2,R} = 0.1$, $b_{2,R} = a_{2,R}/6$,
$w_{2,C_1} = 0.2$, $a_{2,C_1} = 2$, $b_{2,C_1} = a_{2,C_1}/3$
$m_1 = 0.12$, $m_2=0.1$.
In Fig.~4(a): $r=0.06$. In Fig.~4(b): $r=0.126$. In Fig.~4(c): $r=0.164$.

\clearpage
\section{Supplemental Figures}

\begin{figure}[ht!]    
	\centering
	\includegraphics[width=16cm]{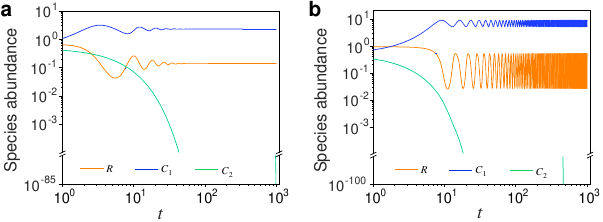}
	\caption{\label{DiscussCEp} 
	Disruption of omnivory in the minimal intraguild predation model~(\ref{P-Psystem2})
	leads to predator extinction. 
	(a-b) Time courses of species abundances showing the extinction of $C_{2}$, 
	simulated from Eq.~(\ref{P-Psystem2}) with omnivory disrupted 
	$\left( \omega_{3}=0, \alpha_{4}=0 \right)$.
	(a): $	\alpha_1 = 0.5, \alpha_2 = 2.5, \alpha_3 = 1.6,
	\omega_1 = 2.3,  \omega_2 = 0.032, 
	m_1 = 0.3, m_2 = 0.2, r = 2.5$.
	(b) : $ \alpha_1 = 2.0, \alpha_2 = 1.6, \alpha_3 = 1.005,
	\omega_1 = 1.95, \omega_2 = 0.038,
	m_1 = 0.25, m_2 = 0.55, r = 6.6$.}   
\end{figure}

\end{document}